\documentclass[a4paper,12pt]{article}
\usepackage[pctex32]{graphics}
\usepackage{amssymb,amsmath}

\begin{document}

\topmargin 0pt \oddsidemargin 0mm
\newcommand{\be}{\begin{equation}}
\newcommand{\ee}{\end{equation}}
\newcommand{\ba}{\begin{eqnarray}}
\newcommand{\ea}{\end{eqnarray}}
\newcommand{\fr}{\frac}
\renewcommand{\thefootnote}{\fnsymbol{footnote}}

\begin{titlepage}

\vspace{5mm}
\begin{center}
{\Large \bf Critical gravity on AdS$_2$ spacetimes }

\vskip .6cm
 \centerline{\large
 Yun Soo Myung$^{1,a}$, Yong-Wan Kim $^{1,b}$,
and Young-Jai Park$^{2,c}$}

\vskip .6cm

{$^{1}$Institute of Basic Science and School of Computer Aided
Science,
\\Inje University, Gimhae 621-749, Korea \\}

{$^{2}$Department of Physics and Department of Service Systems Management and Engineering, \\
Sogang University, Seoul 121-742, Korea}

\end{center}

\begin{center}

\underline{Abstract}
\end{center}
We study the critical gravity in two dimensional AdS (AdS$_2$)
spacetimes, which was obtained from the cosmological topologically
massive gravity (TMG$_\Lambda$) in three dimensions by using the
Kaluza-Klein dimensional reduction. We perform the perturbation
analysis around AdS$_2$, which may correspond to the near-horizon
geometry of the extremal BTZ black hole obtained from the
TMG$_\Lambda$ with identification upon uplifting three dimensions. A
massive  propagating scalar mode $\delta F$ satisfies the
second-order differential equation away from the critical point of
$K=l$, whose solution is given by the Bessel functions. On the other
hand, $\delta F$ satisfies the fourth-order equation at the critical
point. We exactly solve the fourth-order equation, and compare it
with the log-gravity in two dimensions. Consequently, the critical
gravity in two dimensions could not be described by a massless scalar
$\delta F_{\rm ml}$ and its logarithmic partner $\delta F^{\rm
4th}_{\rm log}$. \vspace{5mm}


\vskip 0.8cm

\vspace{15pt} \baselineskip=18pt
\noindent $^a$ysmyung@inje.ac.kr \\
\noindent $^b$ywkim65@gmail.com\\
\noindent $^c$yjpark@sogang.ac.kr

\thispagestyle{empty}
\end{titlepage}

\newpage
\section{Introduction}

The gravitational Chern-Simons (gCS) terms in three dimensional
(3D) Einstein gravity produce a physically propagating massive
graviton~\cite{DJT}. This topologically massive gravity with a
negative cosmological constant $\Lambda=-1/l^2$
(TMG$_\Lambda$~\cite{DCT}) gives us the AdS$_3$
solution~\cite{LSS1}. For the positive Newton's constant $G_3$, a
massive graviton mode carries ghost (negative energy) on the
AdS$_3$. In this sense, the AdS$_3$ is not a stable vacuum. The
opposite case of $G_3<0$ may cure the problem, but it may induce a
negative Deser-Tekin  mass for the BTZ black hole~\cite{DT}. It
seems that there is one way  of  avoiding negative energy by
choosing the chiral (critical)  point of $ K=l$ with the gCS
coupling constant $K$. At this point, a massive graviton becomes a
massless left-moving graviton, which carries no energy. It may be
considered as gauge-artefact. However, the critical point has
raised many questions on physical degrees of freedom
(DOF)~\cite{CDWW1,GJ,GKP,Park,GJJ,CDWW2,Carl,Stro}.

The gCS terms are not invariant under coordinate transformations
though they are conformally invariant~\cite{GIJP,GK}. It is  known
that the 3D Einstein gravity is locally trivial, and thus, does not
have any physically propagating modes. However, all solutions to the
Einstein gravity are also solutions to the TMG$_\Lambda$. Therefore,
it would be better to seek another method to find a propagating
massive mode in  the TMG$_\Lambda$ since it is likely a candidate
for a nontrivial 3D gravity, in addition to the new massive
gravity~\cite{BHT}. To this end, one may introduce a conformal
transformation and then, the Kaluza-Klein reduction  can be  used to
obtain an effective two-dimensional action (2DTMG$_\Lambda$), which
becomes a gauge and coordinate invariant action.  Saboo and
Sen~\cite{SSen,AFM} have used the 2DTMG$_\Lambda$ to derive the
entropy of extremal BTZ black hole~\cite{BTZ} by using the entropy
function formalism (AdS$_2$ attractor equation). When using  the
Achucarro-Ortiz type of dimensional reduction,  it turned out that
there is no propagating massive mode on AdS$_2$
background~\cite{KMP}.

In this work, we will focus on  the chiral point of $K=l$, where a
massive graviton $\psi^M_{mn}$ turned out to be a left-moving
graviton $\psi^L_{mn}$~\cite{LSS1,ssol}.  Grumiller and Johanson
have  introduced a new field $\psi^{\rm
new}_{mn}=\partial_{l/K}\psi^M_{mn}|_{K=l}$ as a logarithmic
parter of $\psi^L_{mn}$~\cite{GJ} based on the logarithmic
conformal field theory (LCFT) with
$c_L=0$~\cite{LCFT3,LCFT4,LCFT5,LCFT6}. However, it was reported
that $\psi^{\rm new}_{mn}$ might not be a physical field at the
chiral point,  since   it belongs to the nonunitary theory. This
is so  because $(\psi^L_{mn},\psi^{\rm new}_{mn})$ become a pair
of dipole ghost fields~\cite{MyungL}.  At this stage, we would
like to mention that the linearized  higher dimensional critical
gravities were recently investigated in the AdS
spacetimes~\cite{hcg}, but the nonunitary  issue of the
log-gravity is not still resolved, indicating that the log-gravity
suffers from the ghost problem.

A few years ago, we have carried out perturbation analysis of the
2DTMG$_\Lambda$ around  AdS$_2$ background~\cite{mkp1}. We have
shown that the dual scalar $\delta F$ of the Maxwell field is a
gauge-invariant  massive mode  propagating in the AdS$_2$
background. Recently, we have studied the critical gravity arisen
from the new massive gravity by investigating quasinormal modes to
check the stability of  the BTZ black hole~\cite{mkp2}.

Hence it is interesting to study the critical gravity arisen from the
2DTMG$_\Lambda$, which shows  a fourth-order differential equation
on AdS$_2$ background.

The organization of our work ia as follows.  In Section 2, we study
the 2DTMG$_\Lambda$, which was obtained from the TMG$_\Lambda$  by
using the Kaluza-Klein dimensional reduction. In Section 3,  we
briefly review the perturbation analysis around AdS$_2$, which may
correspond to the near-horizon geometry of the extremal BTZ black
hole obtained from the TMG$_\Lambda$  with identification upon
uplifting three dimensions.  We find  an explicit solution of  a
physically propagating scalar mode $\delta F$ satisfying  the
second-order differential equation away from the critical point of
$K=l$.    At the critical point, in Section 4, the 2DTMG$_\Lambda$
turns out to be the 2D dilaton gravity including the Maxwell field
obtained from 3D Einstein gravity, which shows that there are no
propagating modes.  We exactly solve the fourth order equation at
the critical point, and compare it with the log-gravity ansatz in
two dimensions. Discussion is given in Section 5.

\section{2DTMG$_\Lambda$}
We start with the action for the TMG$_{\rm \Lambda}$ given
by~\cite{DJT}
\begin{equation} \label{tmg}
I_{\rm TMG_\Lambda}=\frac{1}{16 \pi G_3}\int d^3x\sqrt{-g}\Bigg[R_3
-2\Lambda - \frac{K}{2}\varepsilon^{lmn}\Gamma^p_{~lq}
\Big(\partial_{m}\Gamma^q_{~np}+\frac{2}{3}\Gamma^q_{~mr}\Gamma^r_{~np}\Big)\Bigg],
\end{equation}
where $\varepsilon$ is the tensor  defined by $\epsilon/\sqrt{-g}$
with $\epsilon^{012}=1$. We choose the positive Newton's constant
$G_3>0$ and the negative cosmological constant $\Lambda=-1/l^2$. The
Latin indices of $l,m,n, \cdots$ denote three dimensional tensors.
The $K$-term is called the gCS terms. Here we choose ``$-$" sign in
the front of $K$~\cite{AFM}. Varying this action leads to the
Einstein equation
\begin{equation} \label{eineq}
G_{mn} - KC_{mn}=0,
\end{equation}
where the Einstein tensor  is given by
\begin{equation}
G_{mn}=R_{3mn}-\frac{R_3}{2}g_{mn} -\frac{1}{l^2}g_{mn},
\end{equation}
and the Cotton tensor is defined by
\begin{equation}
C_{mn}= \varepsilon_m~^{pq}\nabla_p
\Big(R_{3qn}-\frac{1}{4}g_{qn}R_3\Big).
\end{equation}
We note that the Cotton tensor $C_{mn}$ vanishes for any solution
to the 3D Einstein gravity, so all solutions of the Einstein
gravity are also solutions of the TMG$_\Lambda$.  Hence, the BTZ
black hole with
 $K=0$~\cite{BTZ} appears as a solution to the full equation (\ref{eineq})
\begin{equation}
ds^2_{\rm BTZ} = -N^2(r) dt^2 + \frac{dr^2}{N^2(r)} + r^2
\Big[d\theta + N^\theta(r) dt\Big]^2 , \label{btzmetric}
\end{equation}
where the squared lapse $N^2(r)$ and the angular shift $N^\theta(r)$
take the forms
\begin{equation}
N^2(r) = -8 G_3 m + \frac{r^2}{l^2} + \frac{16 G_3^2 j^2}{r^2},~~
N^\theta(r) = - \frac{4 G_3 j}{r^2}.\label{def_N}
\end{equation}
Here $m$ and $j$ are the mass and angular momentum of the BTZ black
hole, respectively.

We first make a conformal transformation and then perform
Kaluza-Klein dimensional reduction  by choosing  the
metric~\cite{GIJP,GK}
\begin{equation}
ds^2_{\rm KK}=\phi^2\Big[g_{\mu\nu}(x)dx^\mu
dx^\nu+\Big(d\theta+A_\mu (x)dx^\mu\Big)^2\Big]
\end{equation}
because the gCS terms are invariant under the conformal
transformation.  Here $\theta$ is a coordinate that parameterizes an
$S^1$ with a period $2 \pi l$. Hence, its isometry is factorized as
${\cal G}\times U(1)$. After the ``$\theta$"-integration, the action
(\ref{tmg}) reduces to an effective two-dimensional action called
the 2DTMG$_\Lambda$ as
\begin{eqnarray}\label{2Daction}
 I_{\rm 2DTMG_\Lambda}&=& \frac{l}{8G_3}\int d^2x \sqrt{-g}\left(\phi R+\frac{2}{\phi}g^{\mu\nu}\nabla_\mu\phi\nabla_\nu\phi
     +\frac{2}{l^2}\phi^3-\frac{1}{4}\phi F_{\mu\nu}F^{\mu\nu}\right)\nonumber\\
 &-& \frac{Kl}{32 G_3} \int d^2x \left(R\epsilon^{\mu\nu}F_{\mu\nu}+
     \epsilon^{\mu\nu}F_{\mu\rho}F^{\rho\sigma}F_{\sigma\nu}\right),
\end{eqnarray}
which is {\it our main action to study the critical gravity in two
dimensions.}  Here $R$ is the 2D Ricci scalar with
$R_{\mu\nu}=Rg_{\mu\nu}/2$, and $\phi$ is a dilaton. Also, the
Maxwell field is defined by $F_{\mu\nu}=2\partial_{[\mu}A_{\nu]}$,
and $\epsilon^{\mu\nu}$ is a tensor density. The Greek indices of
$\mu,\nu, \rho, \cdots$ represent two dimensional tensors. Hereafter
we choose $G_3=l/8$ for simplicity. It is again noted that this
action was actively used to derive the entropy of extremal BTZ black
hole by applying the entropy function approach~\cite{SSen,AFM,mkp1}.
Introducing  a dual scalar $F$ of the Maxwell field defined
by~\cite{GIJP,GK}
\begin{equation}
F \equiv -\frac{1}{2\sqrt{-g}}\epsilon^{\mu\nu}F_{\mu\nu},
\end{equation}
equations of motion for $\phi$ and $A_\mu$ are given, respectively,
by
\begin{eqnarray}
 &&
  \label{EOM-phi}
 R+\frac{2}{\phi^2}(\nabla\phi)^2-\frac{4}{\phi}\nabla^2\phi+\frac{6}{l^2}\phi^2+\frac{1}{4}F^2=0,\\
 && \label{EOM-A}
  \epsilon^{\mu\nu}\partial_\nu\left[\phi
  F+\frac{K}{2}(R+3F^2)\right]=0.
\end{eqnarray}
 The equation of motion for the metric $g_{\mu\nu}$ takes
the form
\begin{eqnarray}\label{EOM-g}
 && g_{\mu\nu}\left(\nabla^2\phi-\frac{1}{l^2}\phi^3+\frac{1}{4}\phi F^2-\frac{1}{\phi}(\nabla\phi)^2\right)
    +\frac{2}{\phi}\nabla_{\mu}\phi\nabla_{\nu}\phi-\nabla_{\mu}\nabla_{\nu}\phi
   \nonumber\\
 &&
 +\frac{K}{2}\left[g_{\mu\nu}\left(\nabla^2F+F^3+\frac{1}{2}RF\right)-\nabla_{\mu}\nabla_{\nu}F\right]=0.
\end{eqnarray}
The trace part of Eq. (\ref{EOM-g})
\begin{equation}
 \label{trgEOM}
 \nabla^2\phi-\frac{2}{l^2}\phi^3+\frac{1}{2}\phi F^2
 +K\left(\frac{1}{2}RF+F^3+\frac{1}{2}\nabla^2F\right)=0
\end{equation}
is relevant to our perturbation study.
On the other hand, the traceless part is given by
\begin{equation}
 \label{trlessgEOM}
 g_{\mu\nu}\left(\frac{1}{2}\nabla^2\phi-\frac{1}{\phi}(\nabla\phi)^2\right)
    +\frac{2}{\phi}\nabla_{\mu}\phi\nabla_{\nu}\phi-\nabla_{\mu}\nabla_{\nu}\phi
    +\frac{K}{4}g_{\mu\nu}\nabla^2F-\frac{K}{2}\nabla_{\mu}\nabla_{\nu}F=0
\end{equation}
which may provide a redundant constraint~\cite{KMP}.   Now, we are
in a position to find AdS$_2$ spacetimes as a vacuum solution to
(\ref{EOM-phi}), (\ref{EOM-A}), and (\ref{trgEOM}). In  case of a
constant dilaton, from  (\ref{EOM-phi}) and (\ref{trgEOM}), we have
the condition of a vacuum state
\begin{equation}
(3KF+2\phi)\left(\frac{\phi^2}{l^2}-\frac{1}{4}F^2\right)=0,
\end{equation}
which provides  two distinct relations between $\phi$ and $F$
\begin{eqnarray}
\label{warp} && \phi_{\pm} = \pm\frac{l}{2}F.
\end{eqnarray}
Assuming the line element preserving ${\cal G}=SL(2,R)$ isometry
\begin{equation}\label{ads11}
ds^2_{\rm AdS_2}=\bar{g}_{\mu\nu}dx^\mu
dx^\nu=v\left(-r^2dt^2+\frac{dr^2}{r^2}\right),
\end{equation}
we have the AdS$_2$ spacetimes, which satisfy
\begin{equation} \label{ads2}
\bar{R}=-\frac{2}{v},
~~~\bar{\phi}=u,~~~\bar{F}=\frac{e}{v~}(\bar{F}_{10}=e).
\end{equation}
Here $\bar{F}_{10}=\partial_1 \bar{A}_{0}-\partial_0 \bar{A}_{1}$
with $\bar{A}_{0}=er$ and $\bar{A}_{1}=0$. This background may
correspond to the near-horizon geometry of the extremal BTZ black
hole (NHEB), factorized as AdS$_2 \times S^1$ as \be ds^2_{\rm \pm
NHEB}=\frac{l^2}{4}\Bigg[-r^2dt^2 +\frac{dr^2}{r^2}+(dz\mp
rdt)^2\Bigg], \ee where $v=l^2/4$ and $z=l\theta /|e|$ with the
identification of $z \sim 2\pi l n \frac{l}{|e|}$. Here $n$ is an
integer. As was pointed out in Ref. \cite{BBSS}, the NHEB is a
self-dual orbifold of AdS$_3$. This geometry has a null circle on
its boundary and thus, the dual conformal field theory is a Discrete
Light Cone Quantized (DLCQ) of  CFT$_2$. The kinematics of the DLCQ
show that in a consistent quantum field theory of gravity in these
backgrounds, there is no dynamics in AdS$_2$, which is consistent
with the Kaluza-Klein reduction of the 3D Einstein gravity. However,
the gCS terms in the TMG$_\Lambda$  are odd under parity, and as a
result, the theory shows a single massive propagating degree of
freedom of a given helicity, whereas the other helicity mode remains
massless.  The single massive field is realized as a massive scalar
$\varphi=z^{3/2}h_{zz}$ when using the Poincare coordinates
$x^{\pm}$ and $z$  covering the AdS$_3$
spacetimes~\cite{CDWW1,CDWW2}.  We have  shown that a propagating
massive mode is a dual scalar $\delta F$ of the Maxwell field on a
self dual orbifold of AdS$_3$ (AdS$_2$ background)~\cite{mkp1}.

\section{Perturbation around AdS$_2$}

We briefly review the perturbation around the AdS$_2$ and find the
explicit form of a massive propagating mode. Let us first consider
the perturbation modes of the dilaton, graviton, and dual scalar
around the AdS$_2$ background as
\begin{eqnarray}
 \label{pert1} \phi&=&\bar{\phi}+\varphi, \\
 \label{pert2} g_{\mu\nu}&=&\bar{g}_{\mu\nu}+h_{\mu\nu},~~h_{\mu\nu}=-h\bar{g}_{\mu\nu}\\
 \label{pert3} F&=&\bar{F}(1+\delta F),~~\delta
F=\left(h-\frac{f}{e}\right)
\end{eqnarray}
where the bar variables denote the AdS$_2$ background (\ref{ads11})
and (\ref{ads2}). The Maxwell field has a scalar perturbation $f$
around the background: $F_{10}=\bar{F}_{10}+\delta F_{10}$, where
$\delta F_{10}=-f$. We note that two scalars of $\delta F$ and
$\varphi$ are gauge-invariant quantities in AdS$_2$ spacetimes
although $f$ is not~\cite{mkp1}. Then, considering $\delta R(h)=
\bar{\nabla}^2h-\frac{2}{v}h$, the perturbed equations of motion to
(\ref{EOM-phi}), (\ref{EOM-A}) and (\ref{trgEOM})   are given,
respectively,  by
\begin{eqnarray}
 && \label{pEOMK-phi}
 \bar{\nabla}^2h-\frac{2}{v}h-\frac{4}{u}\bar{\nabla}^2\varphi+\frac{12}{l^2}u\varphi
 +\frac{e^2}{v^2}\delta F=0,\\
 && \label{pEOMK-A}
 \epsilon^{\mu\nu}\partial_\nu\Bigg[\frac{e}{v}\Big(\varphi+u\delta
 F\Big)
 +\frac{K}{2}\Big(\bar{\nabla}^2h-\frac{2}{v}h+\frac{6e^2}{v^2}\delta F\Big)\Bigg]=0, \\
 && \label{pEOMK-g}
 \bar{\nabla}^2\varphi-\frac{6}{l^2}u^2\varphi+\frac{e^2}{2v^2}\varphi+\frac{ue^2}{v^2}\delta F\nonumber\\
 &&
 +K\Bigg(\frac{e}{2v}\bar{\nabla}^2h-\frac{e}{v^2}h+\frac{e(3e^2-v)}{v^3}\delta F
 +\frac{e}{2v}\bar{\nabla}^2\delta F\Bigg)=0.
\end{eqnarray}
Solving (\ref{pEOMK-A})  for  $\delta F$ and inserting it into Eq.
(\ref{pEOMK-g}) leads to
\begin{equation}
\label{pEOMK-g1}
 \left(\bar{\nabla}^2-\frac{2}{v}\right)
 \left(\varphi+\frac{Ke}{2v}\delta F\right)=0
\end{equation}
Also, solving (\ref{pEOMK-A}) for $(\bar{\nabla}^2-\frac{2}{v})h$
and then, inserting it into Eq. (\ref{pEOMK-phi}) arrives at
\begin{equation}
 \label{pEOMK-phi1}
 \left(\bar{\nabla}^2-\frac{2}{v}\right)\varphi
 -\left(\frac{ue}{2vK}+\frac{5}{4v}\right)\Big(\varphi+u\delta
 F\Big)=0.
\end{equation}
Making use of (\ref{pEOMK-g1}) and  (\ref{pEOMK-phi1}),  $\delta F$
and  $\varphi$  satisfy the coupled equation
\begin{equation}
 \label{pEOMK-phi2}
 \left(\bar{\nabla}^2-\frac{2}{v}\right)\delta F
 -\frac{2v}{Ke}\left(\frac{ue}{2vK}+\frac{5}{4v}\right)\Big(\varphi+u\delta F\Big)=0,
\end{equation}
Acting $\left(\bar{\nabla}^2-\frac{2}{v}\right)$ on
(\ref{pEOMK-phi2}), and then eliminating $\varphi$ again by using
(\ref{pEOMK-g1}), one finds {\it the fourth-order equation for
$\delta F$} as follows
\begin{equation}
 \label{pEOMK-phi3}
 \left(\bar{\nabla}^2-\frac{2}{v}\right)\left[\bar{\nabla}^2-\left(\frac{2}{v}+m^2_{\pm}\right)\right]
 \delta F=0,
\end{equation}
for the two AdS$_2$ solutions of $u=\pm \ell e/2v$ in (\ref{warp}).
Here, the mass squared $m^2_{\pm}$ is given by
\begin{equation}
 \label{masspm}
 m^2_{\pm}=\frac{1}{4v}\left(\pm\frac{l}{K}-1\right)\left(5\pm\frac{l}{K}\right).
\end{equation}
Here we stress that  our mass squared is defined differently  from
the Ref.~\cite{mkp1}.
\begin{figure*}[t!]
   \centering
   \includegraphics{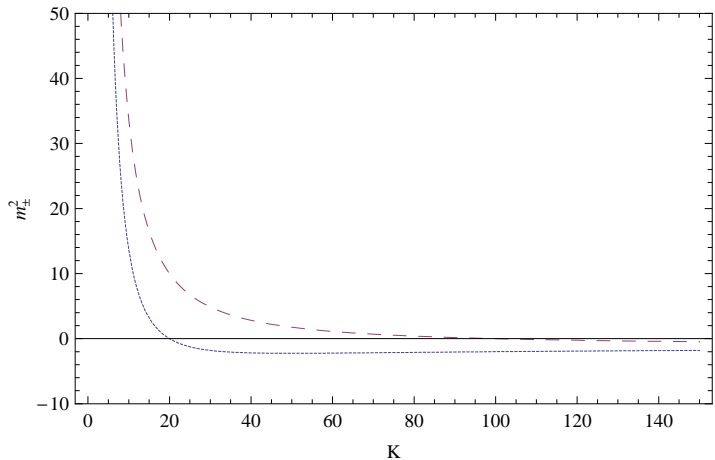}
\caption{Mass $m^2_\pm$ for the AdS$_2$ solution for $l=100$ and
$v=1$: The dotted curve is for the negative  mass squared $m^2_-$,
while the dashed curve is for the positive mass squared $m^2_+=m^2$.
Since the AdS$_2$ solution with a positive charge $q$ is valid for
$K \le l$~\cite{AFM}, the permitted region is $0\le K\le 100$.}
\label{fig.1}
\end{figure*}
For $  0 \le K \le l$, one requires $m^2 \ge0$, which selects
$m^2_+\equiv m^2$ (see Fig. 1). Hereafter we consider this case
only.  For $m^2 \neq 0$, the fourth order equation
(\ref{pEOMK-phi3}) implies two second order equations: one is for a
massless field
\begin{equation}
 \label{pEOMK-ml}
 \left[\bar{\nabla}^2-\frac{2}{v}\right]\delta F=0,
\end{equation}
while the other is for a massive scalar
\begin{equation}
 \label{pEOMK-phi4}
 \left[\bar{\nabla}^2-\left(\frac{2}{v}+m^2\right)\right]\delta F=0,
\end{equation}
In order to solve the massive equation (\ref{pEOMK-phi4}), we
transform the AdS$_2$ metric as
\begin{eqnarray}
 ds^2_{\rm AdS_2} &=& v\left(-r^2dt^2+\frac{dr^2}{r^2}\right) \\
      &\to& -\left(\frac{x^2}{v}\right)  dt^2 + \left(\frac{v}{x^2}\right) dx^2 \\
      &\to& \frac{ds^2}{v}=\left(\frac{1}{x^2_*}\right) \Big[-dt^2 +  dx^2_*\Big],
\end{eqnarray}
in the second line, we used $x=vr$, and in the last line,
$x_*=v/x=1/r$. We note that in the last line, $(t,x_*)$ corresponds
to the Poincar\'e coordinates $(T,y)$ used in Ref.~\cite{SS} to
construct the Hadamard Green function for the Poincar\'e.

Finally,  we wish to  find  a positive  frequency mode for $\delta
F$ as \be \delta F(t,x_*)=e^{-i\omega t} \delta f(x_*). \ee Then,
the second-order equation (\ref{pEOMK-phi4}) becomes
\begin{equation}
\frac{d^2}{dx^2_*}\delta
f+\left[\omega^2-\frac{(m^2v+2)}{x^2_*}\right]\delta f=0,
\end{equation}
whose solution is given by the Bessel functions
\begin{equation}
 \delta f(x_*) = c_1 \sqrt{x_*} J_\nu(\omega x_*) + c_2 \sqrt{x_*} Y_\nu(\omega x_*)
\end{equation}
where $\nu={\sqrt{m^2 v+\frac{9}{4}}}$ satisfying
$\nu^2-1/4=m^2v+2$. Also,  we observe that the event horizon is
located at $r\to0~(x_*\to \infty)$, while the infinity is located at
$r\to \infty~(x_*\to 0)$. In order to have the normalizable
solution, we  choose $c_2=0$ because $ Y_\nu(\omega x_*)$ blows up
at $x_*=0$.

\section{Critical gravity in two dimensions}

At the critical point of $m^2=0~(K=l)$, (\ref{pEOMK-phi3}) becomes
the fourth-order differential equation
\begin{equation} \label{fourtheq}
 \left(\bar{\nabla}^2-\frac{2}{v}\right)^2 \delta F^{\rm 4th}=0.
\end{equation}
In order to solve this equation,  first of all, we observe that the
Bessel function of order $\nu=3/2$ satisfies the second-order
equation for a massless scalar on AdS$_2$ spacetimes as follows:
\begin{eqnarray} \label{seeqq}
 &&  \left(\bar{\nabla}^2-\frac{2}{v}\right) \delta F_{\rm ml}=0,
\end{eqnarray}
whose normalizable solution is given by \be \label{besol} \delta
F_{\rm ml}(t,x_*)=e^{-i\omega t} \delta f_{\rm ml}(x_*) \ee where
\be \label{besol2} \delta f_{\rm ml}(x_*)\simeq \sqrt{x_*}
J_{3/2}(\omega x_*)=\sqrt{\frac{2}{\pi\omega}}\Bigg[-\cos(\omega
x_*)+\frac{\sin(\omega x_*)}{\omega x_*}\Bigg]. \ee

\begin{figure*}[t!]
   \centering
   \includegraphics{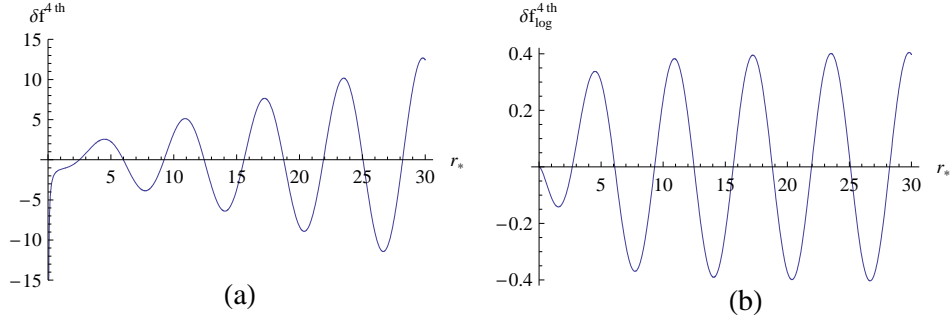}
\caption{Graphs of two functions $\delta f^{\rm 4th}(r_*)$ and the
logarithmic partner $\delta f^{\rm 4th}_{\rm log}(r_*)$ of a
normalizable function $\delta f_{\rm
ml}(r_*)=\sqrt{r_*}J_{3/2}(r_*)$. Although the former is a truly
solution to the fourth-order equation (\ref{feqq}), it shows
singular behavior at infinity of $r\to \infty~(r_*\to 0)$, which may
not be acceptable as a true solution. On the other hand, even though
the logarithmic partner $\delta f^{\rm 4th}_{\rm log}(r_*)$
approaches zero at $r_*=0$, but it is unlikely  a solution to the
fourth-order equation (\ref{feqq}). } \label{fig.2}
\end{figure*}

At this stage, we remind the reader that two equations
(\ref{fourtheq}) and (\ref{seeqq}) with $K=l$ are the same equations
\begin{equation} \label{3dgravity}
 \left(\bar{\nabla}^2-\frac{2}{v}\right)^2 h=0,~~ \left(\bar{\nabla}^2-\frac{2}{v}\right) \varphi=0.
\end{equation}
for the graviton and dilaton as found from the 3D Einstein gravity
with $K=0$~\cite{mkp1}. Here we observe the important
correspondence as \be \delta F_{\rm ml} \leftrightarrow
\varphi,~~~\delta F^{\rm 4th} \leftrightarrow h. \ee   In the 3D
linearized Einstein gravity, one confirms the connection between
dilaton and dual scalar \be \frac{\varphi}{u}=-\delta F. \ee This
means that there are no propagating massive modes at the critical
point, showing apparently that all modes of $h, \varphi,$ and
$\delta F$ from the 3D Einstein gravity are gauge-artefacts.
However, it was proposed that any critical gravity has a new field
on AdS spacetimes. In order to explore this idea on the AdS$_2$
spacetimes, we consider a positive frequency fourth-order field
\be \delta F^{\rm 4th}(t,x_*)=e^{-i\omega t} \delta f^{\rm
4th}(x_*). \ee Then, the fourth order equation (\ref{fourtheq})
takes the form \be \label{feqq}
 \left[\frac{d^2}{dx^2_*}+\left(\omega^2-\frac{2}{x^2_*}\right)\right]^2\delta
f^{\rm 4th}=0. \ee Replacing $\omega x_*=\omega/r$ by $r_*$ and
considering \be\label{4theq} \delta f^{\rm 4th}(r_*)=g(r_*) \delta
f_{\rm ml}(r_*), \ee (\ref{feqq}) reduces to the second-order
equation for $g(r_*)$ as \be \label{geq}
\Big[g''(r_*)-1\Big]=-\frac{2\delta f'_{\rm ml}(r_*)}{\delta f_{\rm
ml}(r_*)}g'(r_*), \ee where the prime $'$ denotes the
differentiation with respect to its argument. Plugging
(\ref{besol2}) into (\ref{geq}) leads to the exact solution for
$g(r_*)$ \be \label{1stg} g(r_*)=C_2+\frac{2C_1
\cos(r_*)+r_*(r_*+2C_1) \sin(r_*)}{2[r_* \cos(r_*)-\sin(r_*)]} \ee
with two undetermined parameters $C_1$ and $C_2$. Hereafter we set
$C_1=-1$ and $C_2=1$ for simplicity. Now, making use of the two
identities
\begin{eqnarray}
  \sin(r_*) &=&\sqrt{\frac{\pi}{2}}\left(\frac{3}{2\sqrt{x}}J_{3/2}(r_*)+\sqrt{r_*}J'_{3/2}(r_*)\right),\nonumber\\
  \cos(r_*)
  &=&\sqrt{\frac{\pi}{2}}\left(\frac{(3-2r^2_*)}{2r^{3/2}_*}J_{3/2}(r_*)+\frac{1}{\sqrt{r_*}}J'_{3/2}(r_*)
  \right),
\end{eqnarray}
we have finally obtained a solution to the fourth-order equation
(\ref{feqq}) as
\begin{equation} \label{ftsol}
  \delta f^{\rm 4th}(r_*) = -
  \left(\frac{3+r^2_*-\frac{1}{2}r^3_*}{2r^{5/2}_*}\right)J_{3/2}(r_*)
  -\left(\frac{1+r^2_*+\frac{1}{2}r^3_*}{r^{3/2}_*}\right)J'_{3/2}(r_*),
  \end{equation}
where $J'_{3/2}(r_*)$ can be expressed in terms of the lower order
Bessel functions as \be
J'_{3/2}(r_*)=\Big(1-\frac{3}{r_*^3}\Big)J_{1/2}+\frac{2}{2r_*^{3/2}}J_{-1/2}.
\ee This shows that $J'_{3/2}(r_*)$ does not contain any singularity
at infinity $r=\infty~(r_*=0)$.   Fig. (\ref{fig.2}a) shows its
behavior on $r_*$ clearly. To see it  more explicitly, $g(r_*)$
takes a series form near $r_*\sim 0~(r\rightarrow\infty)$ \be g(r_*)
\simeq
-\frac{3}{r^3_*}-\frac{9}{5r_*}-\frac{1}{2}+\frac{36}{175}r_*+
\frac{1}{10}r^2_*+\frac{47}{7875}r_*^3+\frac{1}{350}r^4_*\cdots. \ee
Therefore, $\delta f^{\rm 4th}(r_*)$ shows a negative infinity as
\be \label{ddv} -\frac{1}{r_*}~~{\rm as}~~ r_* \to 0 \ee by
observing the first term of \be \delta f^{\rm 4th}
(r_*)=g(r_*)\delta f_{\rm ml}(r_*)\simeq
\sqrt{\frac{2}{\pi}}\Bigg(-\frac{1}{r_*}-\frac{r_*}{2}-\frac{r_*^2}{6}+\frac{r^3_*}{8}+
\frac{r^4_*}{20}-\frac{r^5_*}{144}-\frac{r^6_*}{336}\cdots\Bigg).
\ee For $C_1=-1$ and $C_2=1$, we have a positive infinity of  $
\delta f^{\rm 4th}(r_*) \to \frac{1}{r_*}$ as $r_* \to 0$.

On the other hand, inspired by the log-gravity~\cite{GJ,MyungL}, we
suggest that a solution to the fourth-order equation (\ref{feqq})
may take the form as a logarithmic partner of $\delta F_{\rm
ml}$~\cite{ASS} \be \delta F^{\rm 4th}_{\rm log}(r_*)=e^{-i\omega t}
\delta f^{\rm 4th}_{\rm log}(r_*), \ee where
\begin{eqnarray} \label{losol}
 \delta f^{\rm 4th}_{\rm log}(r_*) &=& \left.\frac{\partial}{\partial
                           m^2}\{\sqrt{r_*}J_\nu(r_*)\}\right|_{m^2=0}\nonumber\\
 &=&        \frac{v\sqrt{r_*}}{3}\Bigg[J_{3/2}(r_*)
            \ln(r_*/2) -\Big(\frac{r_*}{2}\Big)^{3/2} \sum^\infty_{k=0}(-1)^k
            \frac{\psi(5/2+k)}{\Gamma(5/2+k)}\frac{(\frac{1}{4}r^2_*)^k}{k!}\Bigg].
\end{eqnarray}
Here $\Gamma(z)$ is the Gamma function and $\psi(z)$ is a diagamma
function defined by $\psi(z)=\frac{d \ln \Gamma(z)}{dz}$. Fig.
(\ref{fig.2}b) describes $ \delta f^{\rm 4th}_{\rm log}(r_*)$. In
the case of $r_* \to 0$, one has a series form for $\delta f^{\rm
4th}_{\rm log}(r_*)$ as
\begin{eqnarray}
  \delta f^{\rm 4th}_{\rm log}(r_*) &\simeq&  \frac{v}{27}\sqrt{\frac{2}{\pi}}
   \Bigg([-8+3\gamma+3\ln(2r_*)]r_*^2
          -\frac{[-46+15\gamma+15\ln(2r_*)]r_*^4}{50} \nonumber\\
         &&~~~~~~~~~~ +\frac{[-352+105\gamma+105\ln(2r_*)]r_*^6}{9800}+\cdots\Bigg)
\end{eqnarray}
with $\gamma$ the Euler constant.  From this form, we find that
$\delta f^{\rm 4th}_{\rm log}(r_*)$ approaches zero  as  $r_* \to 0$
even though the logarithmic terms are present. Applying the
l'Hospital's rule to $r^n_* \ln(r_*/2)$ with $n \ge1 $
[equivalently, $J_{3/2}(r_*)\ln(r_*/2)]$ as $r_* \to 0$, one finds
immediately that these approach 0.  This shows clearly a different
divergent behavior from (\ref{ddv}). Unfortunately, it is unlikely
that $\delta f^{\rm 4th}_{\rm log}(r_*)$ satisfies the fourth-order
equation (\ref{feqq}). Hence we exclude it as a solution at the
critical point.

Since the solution to the fourth-order solution (\ref{ftsol}) is
singular at $r_* \to0~(r\to \infty)$, it has a problem to be
considered as the normalizable function at infinity.  Hence we
need to care the divergence of $\frac{1}{r_*}$ as $r_* \to 0$
(equivalently, $r$ as $r \to \infty$).

\begin{figure*}[t!]
   \centering
   \includegraphics{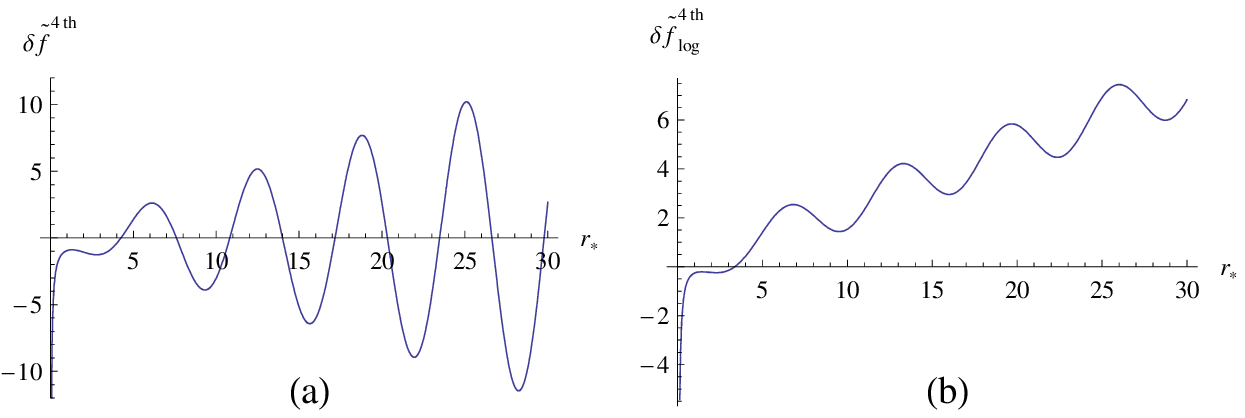}
\caption{Graphs of two functions $\delta \tilde{f}^{\rm 4th}(r_*)$
and the logarithmic partner $\delta \tilde{f}^{\rm 4th}_{\rm
log}(r_*)$ of a nonnormalizable function $\delta \tilde{f}_{\rm
ml}(r_*)=\sqrt{r_*}Y_{3/2}(r_*)$. Although the former is a  solution
to the fourth-order equation (\ref{feqq}), it shows singular
behavior at infinity of $r\to \infty~(r_*\to 0)$, which may not be
acceptable as a true solution.  On the other hand,  the logarithmic
partner $\delta \tilde{f}^{\rm 4th}_{\rm log}(r_*)$ shows a singular
behavior at $r_*=0$ and  it is unlikely a solution to the
fourth-order equation (\ref{feqq}).} \label{fig.3}
\end{figure*}

On the other hand, we may choose the second kind of Bessel
function $Y_{3/2}$ as a solution of the second-order equation for
a massless scalar on the AdS$_2$ spacetimes  even it belongs to
the nonnormalizable function at infinity as
\begin{equation}
 \delta \tilde{f}_{ml}(x_*) \simeq \sqrt{r_*}Y_{3/2}(\omega x_*)
 =\sqrt{\frac{2}{\pi\omega}}
  \Bigg[-\sin(\omega x_*) -\frac{\cos(\omega x_*)}{\omega x_*} \Bigg].
\end{equation}
After replacing $\omega x_*=\omega/r$ by $r_*$, and solving
(\ref{geq}), we have
\begin{equation}\label{geq2}
 \tilde{g}(r_*)=\tilde{C}_2+\frac{ \tilde{C}_1 \sin(r_*)-r_*(2r_*+\tilde{C}_1) \cos(r_*)}{4[\cos(r_*)+r_* \sin(r_*)]}
\end{equation}
instead of (\ref{1stg}). Near $r_*\sim 0$, we have a regular behavior as
\begin{equation}
 \tilde{g}(r_*)\simeq
 1-\frac{r^2_*}{2}+\frac{r^3_*}{12}+\frac{r^4_*}{2}-\frac{r^5_*}{20}-\frac{r^6_*}{3}\cdots,
\end{equation}
with $\tilde{C}_1=\tilde{C}_2=1$. Making use of the two identities
\begin{eqnarray}
   \sin(r_*) &=&\sqrt{\frac{\pi}{2}}\Bigg[\frac{(3-2r^2_*)}{2r^{3/2}_*}Y_{3/2}(r_*)+\frac{1}{\sqrt{r_*}}Y'_{3/2}(r_*)\Bigg],\nonumber\\
  \cos(r_*) &=& -\sqrt{\frac{\pi}{2}}\Bigg[\frac{3}{2\sqrt{r_*}}Y_{3/2}(r_*)+\sqrt{r_*}Y'_{3/2}(r_*)\Bigg],
\end{eqnarray}
we find {\it another solution to the fourth-order equation (\ref{feqq}) as }
\begin{equation}
 \delta \tilde{f}^{\rm 4th}(r_*) =
 \left(\frac{-3-r^2_*+2r^3_*}{8r^{5/2}_*}\right)Y_{3/2}(r_*)
                          -\left(\frac{1+r^2_*+2r^3_*}{4r^{3/2}_*}\right)Y'_{3/2}(r_*).
\end{equation}
However, Fig. (\ref{fig.3}a) shows its singular behavior as  $r_*
\to 0$, too.  Near $r_*\sim 0~(r\rightarrow\infty)$, one has a
divergence of $-\frac{1}{0}$ as
\begin{equation}
 \delta \tilde{f}^{\rm 4th}(r_*)=\tilde{g}(r_*)\delta \tilde{f}_{\rm ml}(r_*)
           \simeq \sqrt{\frac{2}{\pi}}\left(-\frac{1}{r_*}-\frac{r^2_*}{12}-\frac{r^3_*}{8}
           +\frac{r^4_*}{120}+\frac{r^5_*}{72}-\frac{r^6_*}{3360}\cdots\right).
\end{equation}

Finally, introducing  the log-gravity, a suggested solution as a
logarithmic partner of $\delta \tilde{f}_{ml}$ takes the form~\cite{ASS}
of \be \delta \tilde{F}^{\rm 4th}_{\rm
log}(r_*)=e^{-i\omega t} \delta \tilde{f}^{\rm 4th}_{\rm log}(r_*),
\ee where
\begin{eqnarray}
 &&  \delta \tilde{f}^{\rm 4th}_{\rm log}(r_*) = \left.\frac{\partial}{\partial
                           m^2}\{\sqrt{r_*}Y_\nu(r_*)\}\right|_{m^2=0}\nonumber\\
 && = \frac{v\sqrt{r_*}}{3}\left[\ \cot[3\pi/2]\left(
                                       J_{3/2}(r_*) \ln(r_*/2)
            -\Big(\frac{r_*}{2}\Big)^{3/2} \sum^\infty_{k=0}(-1)^k
            \frac{\psi(5/2+k)}{\Gamma(5/2+k)}\frac{(\frac{1}{4}r^2_*)^k}{k!}-\pi Y_{3/2}(r_*)\right)\right.
            \nonumber\\
        && ~~~~ + \csc[3\pi/2]\left(J_{-3/2}(r_*) \ln(r_*/2)
            -\Big(\frac{r_*}{2}\Big)^{-3/2} \sum^\infty_{k=0}(-1)^k
            \frac{\psi(-1/2+k)}{\Gamma(-1/2+k)}\frac{(\frac{1}{4}r^2_*)^k}{k!}\right)
        -\pi J_{3/2}(r_*)\Bigg] \nonumber\\
 && =
 \frac{v}{3}\left(\frac{3a}{\sqrt{2\pi}}r_*-\pi\sqrt{r_*}J_{3/2}(r_*)-\ln(r_*/2)Y_{3/2}(r_*)\right)
\end{eqnarray}
with $a=0.616108$. Fig. (\ref{fig.3}b) indicates  $\delta
\tilde{f}^{\rm 4th}_{\rm log}(r_*)$ is singular at $r_*=0$. To show
it explicitly, one finds a series expansion of $\tilde{f}^{\rm
4th}_{\rm log}(r_*)$ as
\begin{eqnarray}
  \tilde{f}^{\rm 4th}_{\rm log}(r_*) &\simeq& v \sqrt{\frac{2}{\pi}}
    \left[
    \left(\frac{1}{3r_*}+\frac{1}{6}r_*
              -\frac{1}{24}r^3_*+\frac{1}{432}r^5_*\right)\ln(r_*/2)+\left(\frac{a}{2}r_* -\frac{\pi}{9}r^2_*
    +\frac{\pi}{90}r^4_*\right)+\cdots\right].\nonumber\\
\end{eqnarray}
Here we note that  the first term in $\ln(r_*/2)$ shows a singular
behavior as $r_* \to 0$, while the remaining terms makes a finite
graph as an oscillatory  increasing function  for large $r_*$.

\section{Discussions}

We have studied the critical gravity in AdS$_2$ spacetimes, which
was obtained from the topologically massive gravity  in three
dimensions by using  the Kaluza-Klein dimensional reduction.  We
have performed the perturbation analysis around the AdS$_2$, which
corresponds to the near-horizon geometry of the extremal BTZ black
hole obtained from the topological massive gravity with
identification upon uplifting three dimensions. A physically
massive scalar mode $\delta F$ satisfies the second-order
differential equation away from the critical point of $K=l$, while
it satisfies the fourth-order equation at the critical point. At
the critical point, the 2DTMG$_{\rm \Lambda}$ turns out to be the
2D dilaton gravity including the Maxwell field obtained from the
3D Einstein gravity, which shows implicitly  that there are no
propagating modes.

Based on that the critical gravity has a new field in AdS
spacetimes,  we have exactly solved the fourth-order equation, and
compared it with the log-gravity ansatz in two dimensions.  The
critical gravity is described by $\delta f^{\rm 4th}$
(\ref{ftsol}) precisely, which, however, it becomes divergent
linearly ($r \to \infty$) as the infinity of $r_*=0~(r=\infty)$ is
approached. This means that the solution to the fourth-order
equation is not a precisely normalizable function and thus, it
requires to introduce an appropriate boundary condition which
accommodates a linear divergence.

More importantly, it has turned out that the critical gravity could not
be described by the massless scalar  $\delta f_{\rm ml}$ and its
logarithmic partner $\delta f^{\rm 4th}_{\rm log}$ (\ref{losol}),
which approaches zero as $r_* \to 0$. This is so because  $\delta
f^{\rm 4th}_{\rm log}$ unlikely satisfies the fourth-order equation.

Finally, we would like to comment that  the linearized  higher
dimensional critical gravities were widely  investigated in the
AdS spacetimes~\cite{hcg} but  the non-unitarity issue of the
log-gravity is not still resolved, indicating that any log-gravity
suffers from the ghost problem.   Furthermore, the critical
gravity on the Schwarzschild-AdS black hole has suffered from the
ghost problem when the cross term $E_{\rm cross}$ is
non-vanishing~\cite{LLL}.

 \vskip 0.5cm

\section*{Acknowledgements}
Y.S. Myung and Y.-W. Kim were supported by Basic
Science Research Program through the National Research Foundation
(KRF) of Korea funded by the Ministry of Education, Science and
Technology (2010-0028080). Y.-J. Park was supported by World Class
University program funded by the Ministry of Education, Science and
Technology through the National Research Foundation of Korea(No.
R31-20002). Y. S. Myung and Y.-J. Park were
also supported by the National Research Foundation of Korea (NRF)
grant funded by the Korea government (MEST) through the Center for
Quantum Spacetime (CQUeST) of Sogang University with grant number
2005-0049409.

\end{document}